\documentclass[aps,pra,twocolumn,superscriptaddress,longbibliography]{revtex4-2}

\usepackage{amsthm,amsmath,amsfonts}
\usepackage{longtable,tabularx,subfigure,xcolor,color}
\usepackage{siunitx}
\usepackage{graphicx}
\usepackage{dcolumn}
\usepackage{bm,bbm,bbold}
\usepackage{float}
\usepackage{multirow}
\usepackage{comment}
\usepackage{tabularx}
\usepackage[ruled,vlined]{algorithm2e}
\usepackage{hyperref}

\newtheorem{lemm}{Lemma}

\newtheorem{thrm}{Theorem}

\newcommand{\Tr}{\mathrm{Tr}}

\newcommand{\bra}[1]{\mbox{$\langle #1 |$}}
\newcommand{\ket}[1]{\mbox{$| #1 \rangle$}}

\newcommand{\CC}{\mathbb{C}}

\begin{document}

\title{Quantifying Electron Entanglement Faithfully}

\author{Lexin Ding}
\affiliation{Faculty of Physics, Arnold Sommerfeld Centre for Theoretical Physics (ASC),\\Ludwig-Maximilians-Universit{\"a}t M{\"u}nchen, Theresienstr.~37, 80333 M{\"u}nchen, Germany}
\affiliation{Munich Center for Quantum Science and Technology (MCQST), Schellingstrasse 4, 80799 M{\"u}nchen, Germany}

\author{Zolt\'an Zimbor{\'a}s}
\affiliation{Theoretical Physics Department, Wigner Research Centre for Physics, P.O.Box 49 H-1525, Budapest, Hungary}
\affiliation{Algorithmiq Ltd., Kanavakatu 3C, FI-00160 Helsinki, Finland}
\affiliation{Eötvös Lorán University, Pázmány Péter sétány. 1/C, 1117 Budapest, Hungary}

\author{Christian Schilling}
\email{c.schilling@physik.uni-muenchen.de}
\affiliation{Faculty of Physics, Arnold Sommerfeld Centre for Theoretical Physics (ASC),\\Ludwig-Maximilians-Universit{\"a}t M{\"u}nchen, Theresienstr.~37, 80333 M{\"u}nchen, Germany}
\affiliation{Munich Center for Quantum Science and Technology (MCQST), Schellingstrasse 4, 80799 M{\"u}nchen, Germany}

\date{\today}

\begin{abstract}
Entanglement is one of the most fascinating concepts of modern physics. In striking contrast to its abstract, mathematical foundation, its practical side is, however, remarkably underdeveloped. Even for systems of just two orbitals or sites no faithful entanglement measure is known yet. By exploiting the spin symmetries of realistic many-electron systems, we succeed in deriving a closed formula for the relative entropy of entanglement between electron orbitals. Its broad applicability in the quantum sciences is demonstrated: (i) in light of the second quantum revolution, it quantifies the true physical entanglement by incorporating the crucial fermionic superselection rule (ii) an analytic description of the long-distance entanglement in free electron chains is found, refining Kohn's locality principle (iii) the bond-order wave phase in the extended Hubbard model can be confirmed, and (iv) the quantum complexity of common molecular bonding structures could be marginalized through orbital transformations, thus rationalizing zero-seniority wave function ansatzes.
\end{abstract}

\maketitle
\section{Introduction}
Entanglement is one of the key concepts in the quantum information sciences, particularly due to its significance
for information processing tasks \cite{ekert1991quantum,bennett1992experimental,bennett1992communication,bennett1993teleporting,mattle1996dense,bouwmeester1997experimental}. For example, with the aid of shared entanglement, one can teleport a quantum state without any knowledge of the state itself \cite{bennett1993teleporting,bouwmeester1997experimental}, encode an amount of information beyond classical capacity \cite{mattle1996dense}, or even set up a communication channel where eavesdropping is impossible \cite{bennett2020quantum}.
With such applications in mind, a comprehensive foundation for entanglement has been developed in quantum information theory:
Axioms for entanglement measures were established and formal definitions for measures were suggested such as the ``entanglement of formation'' \cite{bennett1996mixed}, ``entanglement cost'' \cite{hayden2001asymptotic} and ``entanglement of distillation'' \cite{rains1999rigorous}. In particular, enormous effort has been dedicated towards understanding their operational meaning for information processing \cite{horodecki2009quantum,plenio2014introduction}.

In striking contrast to the rather abstract foundation of entanglement, its practical side is remarkable underdeveloped. Even for systems of just two electrons on two lattice sites no closed formula is known for measuring (faithfully) entanglement.
At first sight, this claim seems to be surprising given the
huge body of literature emphasizing the significance of entanglement, e.g., for quantum phase transitions \cite{osborne2002entanglement,osterloh2002scaling,vidal2003entanglement}, topological order \cite{kitaev2006topological,levin2006detecting}, chemical bonding \cite{boguslawski2013orbital,szalay2017correlation}
and the implementation of numerical methods \cite{white1992density,legeza2003optimizing,schollwock2011density,stein2016automated}. Yet, it is crucial to notice that all these works had to make significant concessions, restricting either their scope or conclusiveness. For instance, since the simple von Neumann entropy refers to pure states, it can quantify the entanglement of a subsystem $A$
\emph{only} with the \emph{entire} complementary system, i.e., the universe. For the physically relevant case of two \emph{arbitrary} subsystems $A,B$ --- their joint state $\rho_{AB}$ thus being mixed --- there is no practically feasible alternative to the so-called ``entanglement negativity''. The latter's deficiency, to vanish for some entangled states (non-faithfulness), however, raises doubts about its usefulness for investigating the true role of entanglement in physics, chemistry, and materials science.
A systematic way to avoid all these fundamental deficiencies would be to resort to the relative entropy of entanglement which measures the minimal ``distance'' $S(\rho_{AB}||\sigma^\ast)$ of a quantum state $\rho_{AB}$ to the convex set of unentangled/separable states \cite{vedral1998entanglement}. Intriguingly, the problem of determining for a given boundary point $\sigma^\ast$ of that convex set all $\rho_{AB}$ for which $\sigma^\ast$ is the closest separable state has been solved \cite{miranowicz2008closed,friedland2011explicit}.
Yet, since no efficient description of the boundary of the set of separable states is known, the solution to this inverse problem
will not simplify our task of calculating the relative entropy of entanglement.
Accordingly, the need and the challenging character of deriving a corresponding closed formula can hardly be overestimated.
Even for the case of just two qubits, with Hilbert space $\CC^2 \otimes \CC^2$,
this problem is listed as one of the long-standing problems in quantum information theory \cite{krueger2005some}. The application in many-electron systems would even require a solution for the setting $\CC^4 \otimes \CC^4$ since
the Fock space of a single orbital or lattice site is four dimensional, spanned by
$\{|0\rangle, |\!\uparrow\rangle, |\!\downarrow\rangle, |\!\uparrow\downarrow\rangle\}$.

It will be the main achievement of our work to derive a \emph{closed formula} for exactly this setting, quantifying the entanglement between any two electronic orbitals. Our derivation necessitates a number of non-trivial ingredients such as the superselection rule, the formalism for taking into account local and non-local symmetries and a compact description of the set of unentangled states.
Most importantly, this anticipated key result --- the closed formula \eqref{eqn:rel_ent_formula} --- is information theoretical complete in the sense that the entire information about ground states is encoded in their reduced density matrices of the setting $\CC^4 \otimes \CC^4$. For instance, in lattice models with hopping and interaction restricted to nearest neighbors, their ground state properties are uniquely described by the two-site reduced density matrices $\rho_{i,i+1}$.
This generalizes to arbitrary continuous systems with pair interaction \cite{Col63,Mazz12,Mazz16}, by replacing $\rho_{i,i+1}$ by $\rho_{\vec{x},\vec{y}}$. In turn, this constitutes the prominent \emph{Coulson Challenge} \cite{Col00}:
Finding an efficient description of the corresponding set of $N$-representable reduced density matrices would allow one to solve \emph{efficiently} the ground state problem for any quantum system. 

\section{Notation and Concepts}
\begin{figure}[t]
\centering
\includegraphics[scale=0.38]{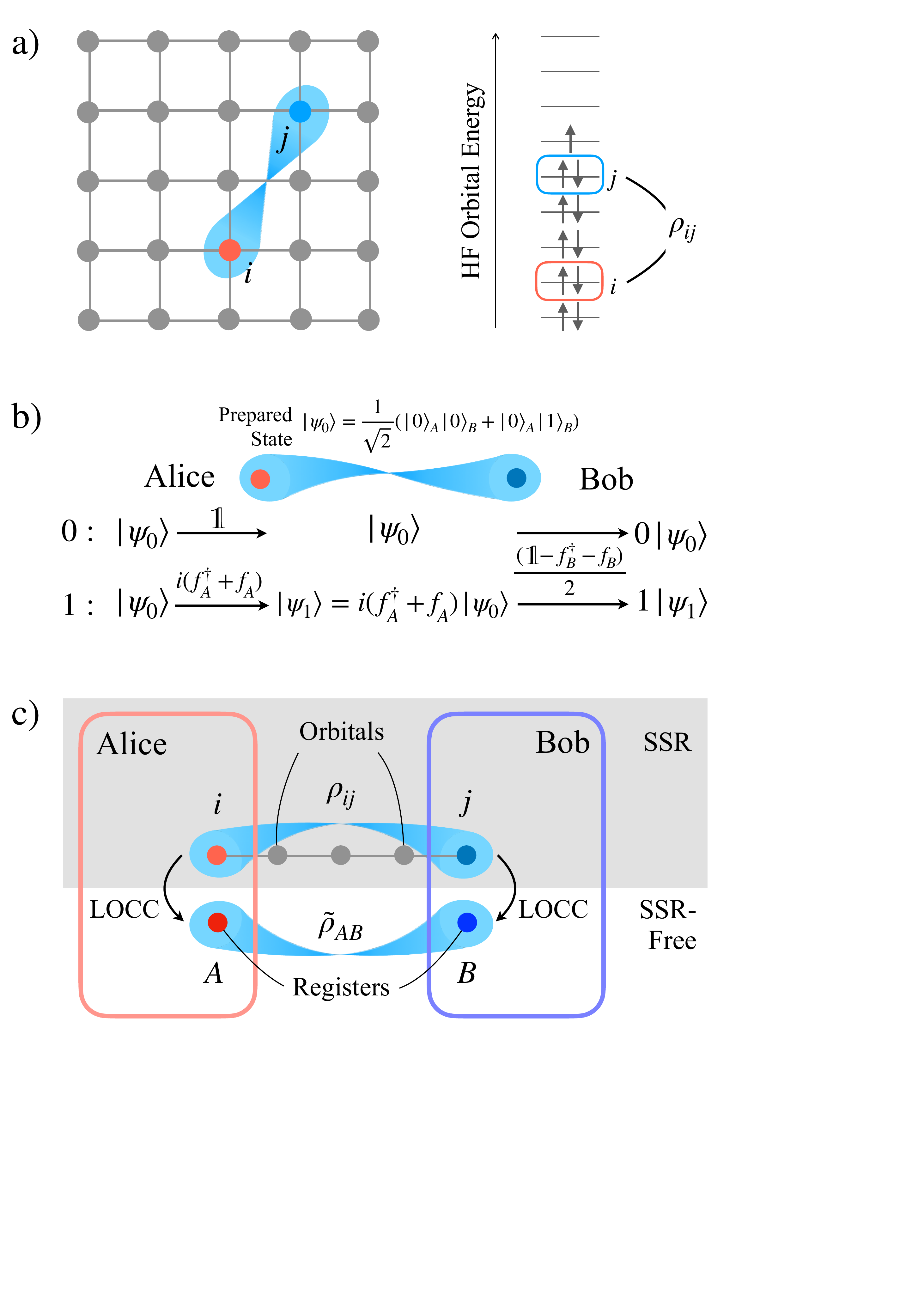}
\caption{a) A 2D Fermi-Hubbard lattice (left) and a set of Hartree-Fock (HF) molecular orbitals ordered with increasing orbital energy (right). b) Alice and Bob performing superluminal signalling of one-bit information with two spinless fermion modes in a world free of superselection rules (SSR). (See text for more information.) c) Alice and Bob transfer the state $\rho_{ij}$ on two physical orbitals $i$ and $j$, to a state $\tilde{\rho}_{AB}$ on the local quantum registers $A$ and $B$ via LOCC.}
\label{fig:resource}
\end{figure}

The concept of entanglement refers to a notion of subsystems. But how could the latter be established in the context of \emph{fermionic} quantum systems? Clearly, due to their indistinguishable nature, individual fermions do not constitute proper subsystems. A meaningful and particularly appealing partition, however, emerges within the framework of ``2nd quantization'', namely by dividing the set of lattice sites or molecular orbitals into disjoint subsets (see Figure \ref{fig:resource}a).

In such fermionic settings, the concept of entanglement is well understood from a conceptual and mathematical point of view \cite{zanardi2002quantum,schuch2004quantum,banuls2007entanglement,friis2013fermionic,eisler2015gaussian,Eisert18,Spee18}, and its application with an emphasis on information processing tasks has become an active field of research \cite{boguslawski2015orbital,gigena2015entanglement,franco2016quantum,franco2018indistinguishability,Pachos18free,ding2020correlation,ding2020concept,morris2020entanglement,benatti2020entanglement,debarba2020teleporting,olofsson2020quantum,aoto2020calculating,galler2021orbital,pusuluk2021classical,faba2021two,faba2021correlation,faba2022Lipkin,sperling2022entanglement,ding2022quantum}. Before considering general fermionic systems, we introduce and explain various relevant concepts first in the simplest possible setting of just \emph{two} spinless fermionic sites, orbitals or modes $A, B$. By referring to those two modes $A,B$ we automatically established a notion of subsystems \cite{zanardi2001virtual}. From a more mathematical point of view, this means to define an unambiguous starting point for our quantum informational analysis by splitting the corresponding two-dimensional one-fermion Hilbert space $\mathcal{H}^{(1)}\cong \mathbb{C}^2$ into two orthogonal subspaces, $\mathcal{H}^{(1)} = \mathcal{H}^{(1)}_A \oplus \mathcal{H}^{(1)}_B$. Each mode can be either empty or occupied and the configuration states $|n_A,n_B\rangle$, where $n_{A,B} = 0,1$, span the corresponding total Fock space accordingly. Finally, by referring to the formal splitting $|n_A, n_B\rangle \mapsto |n_A\rangle \otimes |n_B\rangle$ we can decompose the underlying Fock space, $\mathcal{F}[\mathcal{H}^{(1)}] \cong \mathcal{F}[\mathcal{H}_A^{(1)}]\otimes \mathcal{F}[\mathcal{H}_B^{(1)}]$, and reveal the sought-after tensor product structure of our composite quantum system $AB$.

Before we discuss the entanglement between the two modes, we must incorporate a final fermionic ingredient, a crucial superselection rule \cite{wick1970superselection}. Namely, nature does not allow for every possible operation on fermionic subsystems: local observables must preserve local particle numbers. In other words, observables on mode $A$ must take the form $\hat{O}_A = \lambda_0 |0_A\rangle\langle 0_A| + \lambda_1 |1_A\rangle\langle1_A|$, and likewise for mode $B$. The direct consequence is that only the diagonal blocks of the quantum state $\rho$ with fixed local particle numbers $N_{\!A}, N_{\!B}$ can be observed in reality. This then means that $\rho$ can never be distinguished through local observables from its so-called superselected variant $\tilde{\rho}$ \cite{bartlett2003entanglement,banuls2007entanglement}. Since $A$ and $B$ are both spinless modes, the superselection works quite simply. We demonstrate here with a general state $\rho$ (in the ordered basis $|0,0\rangle,|0,1\rangle,|1,0\rangle,|1,1\rangle$) the effect of the particle number superselection rule
\begin{equation}
\rho =\left[\begin{smallmatrix}
\rho_{11} & \rho_{12} & \rho_{13} & \rho_{14}
\\
\rho_{21} & \rho_{22} & \rho_{23} & \rho_{24}
\\
\rho_{31} & \rho_{32} & \rho_{33} & \rho_{34}
\\
\rho_{41} & \rho_{42} & \rho_{43} & \rho_{44}
\end{smallmatrix}\right]\,\xrightarrow{\mathrm{SSR}}\: \tilde{\rho} =\left[\begin{smallmatrix}
\rho_{11} & 0 & 0 & 0
\\
0 & \rho_{22} & 0 & 0
\\
0 & 0 & \rho_{33} & 0
\\
0 & 0 & 0 & \rho_{44}
\end{smallmatrix}\right]\!.
\end{equation}

The generalization of various concepts to arbitrary fermionic systems is straightforward.
To introduce a notion of subsystems, one just divides some chosen orthonormal basis of the underlying (higher-dimensional) one-particle Hilbert space $\mathcal{H}^{(1)}$ into two disjoint subsets $\{\ket{\varphi_i}\}_{i=1}^k$, $\{\ket{\varphi_i}\}_{i=k+1}^d$ whose states then span $\mathcal{H}^{(1)}_A$ and $\mathcal{H}^{(1)}_B$, respectively. Using the corresponding occupation number representation, this leads according to $\ket{n_1,\ldots,n_d}\mapsto \ket{n_1,\ldots,n_k}\otimes \ket{n_{k+1},\ldots,n_d} $ to  a tensor product structure
\begin{equation}
\mathcal{F}[\mathcal{H}^{(1)}] \cong \mathcal{F}[\mathcal{H}^{(1)}_{A}] \otimes \mathcal{F}[\mathcal{H}^{(1)}_{B}].
\end{equation}
On this level, the superselection can be formalized as
\begin{equation}
\rho \mapsto \tilde{\rho} = \sum_{N_{\!A}, N_{\!B} \geq 0} P_{N_{\!A}}\! \otimes \! P_{N_{\!B}} \, \rho \, P_{N_{\!A}}\! \otimes\! P_{N_{\!B}}, \label{eqn:tilde}
\end{equation}
where $P_N$ is the projection onto the $N$-particle sector of the respective Fock space $\mathcal{F}[\mathcal{H}^{(1)}_{A/B}]$.


\section{Superselection Rules and Accessible Entanglement}
The important restriction $\rho \mapsto \tilde{\rho}$ is urged by the fundamental laws of physics. Violating the superselection rule (SSR) would lead to dire scenarios where superluminal signalling becomes possible \cite{johansson2016comment,ding2020concept}(Figure \ref{fig:resource}b), in striking contradiction to the law of special relativity. To explain this key aspect, let us perform a Gedankenexperiment by assuming that the SSR could be violated. Suppose Alice and Bob each holds a spinless fermionic mode $A$ and $B$, respectively. Each mode can either be empty or occupied, resulting in two orthogonal local basis states $|0\rangle_{A/B}$ and $|1\rangle_{A/B} = f^{\dagger}_{A/B} |0\rangle_{A/B}$, where $f^{\dagger}_{A/B}$ denotes the respective fermionic creation operator. The state shared between Alice and Bob could be prepared as $|\Psi\rangle_{AB} = \frac{1}{\sqrt{2}}(|0\rangle_A \otimes |0\rangle_B + |0\rangle_A \otimes |1\rangle_B)$ by applying $\frac{1}{\sqrt{2}}(\openone + f^\dagger_{B})$ to the vacuum.  When the two parties are far away, Alice could then send an instant bit of information ($0$ or $1$) to Bob with the following actions: if Alice wishes to communicate a bit ``$0$", she does nothing (applies $\openone$ to her mode); If she wishes to send ``$1$", she then performs the local unitary operation $U_A = i(f^\dagger_A - f^{\phantom{\dagger}}_A)$. Remarkably, when Bob measures the local observable $\hat{O}_B = \frac{1}{2}(\openone - f^\dagger_B - f^{\phantom{\dagger}}_B)$ the outcome would be definite. To be more specific, one easily verifies that in either case Bob's local state would be an eigenstate of $\hat{O}_B$, and its eigenvalue would coincide with the value of Alice's message.

Clearly, because of the SSR, such paradoxical scenarios would never occur. Instead, this Gedankenexperiment highlights in the most compelling way that the entanglement shared between Alice and Bob was never physical to begin with. Only the entanglement in accordance with SSR is of physical relevance and can be extracted and utilized for quantum information processing tasks. From a practical point of view, the useful entanglement in a fermionic quantum state $\rho$ is that which can be transferred via local operations and classical communications (LOCC) to a new state on a suitable quantum register, on which there is no restriction of SSR. After the transferring process, the state left on the quantum register is precisely $\tilde{\rho}$ \cite{vaccaro2003identical}, whose entanglement can be manipulated with the unrestricted local operations (see Figure \ref{fig:resource}c). Therefore taking into account SSR when quantifying fermionic entanglement is not only the theoretically accurate way, but also of serious practical relevance in the on-going second quantum revolution.


Last but not least, the relative entropy of entanglement $E(\rho)$ is defined \cite{vedral1997quantifying} as the minimal relative entropy
\begin{equation}
S(\rho||\sigma) \equiv \Tr[\rho(\log(\rho)-\log(\sigma))]
\end{equation}
of $\rho$ relative to the set of separable (i.e., unentangled) states.
As it has been motivated above and explained in detail in \cite{bartlett2003entanglement,banuls2007entanglement}, the effect of the SSR can be transferred into the quantum state $\tilde{\rho}$, leading to
\begin{equation}\label{eqn:rel_ent}
    E(\rho) = \min_{\sigma \in \mathcal{D}_{sep}} S(\tilde{\rho}|| \sigma),
\end{equation}
with $\mathcal{D}_{sep}$ being the set of all classical mixtures of product states $\rho_A \otimes \rho_B$.

\section{Derivation of closed formula}

We apply now these general concepts to our context. For this, we first observe that the one-electron Hilbert space factorizes into an orbital part and a spin part, $\mathcal{H}^{(1)}= \mathcal{H}_l^{(1)} \otimes \mathcal{H}_s^{(1)}$, where the latter follows as $\mathcal{H}_s^{(1)} \cong \CC^2$. Accordingly, any decomposition of the orbital part $\mathcal{H}_l^{(1)}$ implies a corresponding decomposition of the full one-particle Hilbert space $\mathcal{H}^{(1)}$.

The notion of entanglement between two orbitals $\ket{\varphi_1}, \ket{\varphi_2} \in \mathcal{H}_l^{(1)}$ thus refers to the splitting
\begin{equation}
\mathcal{F}[\mathcal{H}^{(1)}] \cong \mathcal{F}[\mathcal{H}_A^{(1)}]\otimes \mathcal{F}[\mathcal{H}_B^{(1)}]\otimes \mathcal{F}[\mathcal{H}_C^{(1)}],
\end{equation}
where $\mathcal{H}_{A/B}^{(1)}$ is spanned by the two spin-orbitals $\ket{\varphi_{1/2}}\otimes \ket{\!\uparrow}$, $\ket{\varphi_{1/2}}\otimes \ket{\!\downarrow}$ and  $\mathcal{H}_{C}^{(1)}$ by all the remaining ones.
The orbital-entanglement between $\ket{\varphi_1}$ and $\ket{\varphi_2}$ then follows by first
tracing out the complementary system $C$ and then calculating the relative entropy of entanglement \eqref{eqn:rel_ent}
of the reduced state $\rho\equiv \rho_{AB}$. For example, if we wish to study the entanglement between two sites $i$ and $j$ in a lattice system, or the entanglement between two Hartree-Fock orbitals $i$ and $j$, we must first trace out all other site/orbital degrees of freedom except for site/orbital $i$ and $j$ (see Figure \ref{fig:resource}a). The sought-after entanglement between the two sites/orbitals can then be calculated from the resulting reduced state $\rho_{ij}$.

Since the Fock space of one orbital is spanned by the states \mbox{$\{|0\rangle, |\!\uparrow\rangle, |\!\downarrow\rangle, |\!\uparrow\downarrow\rangle\}$}, where $\ket{0}$ denotes the vacuum state, the mathematical setting underlying our following derivation is $\mathbb{C}^4 \otimes \mathbb{C}^4$. Moreover, since $\rho,\sigma$ in \eqref{eqn:rel_ent} are density operators on a 16-dimensional Hilbert space, the corresponding minimization problem \eqref{eqn:rel_ent} would in principle involve  $16\times16-1=255$ real parameters. Even worse,  no efficient description of the convex set $\mathcal{D}_{sep}$ is known and our task of deriving a closed formula seems to be hopeless. In contrast to generic states, two-orbital reduced density operators $\rho$ of realistic many-electron quantum states exhibit, however, a number of simplifying symmetries. Referring to the most relevant scenarios in quantum chemistry and solid state physics, we assume in the following
\begin{equation}
[\rho,\hat{N}]=[\rho,\hat{S}^z]=[\rho, \hat{\vec{S}}^2]=0,
\end{equation}
where $\hat{N}, \hat{S}^z, \hat{\vec{S}}^2$ denote the particle number and the spin operators of the two-orbital subsystem. The first two symmetries of $\rho$ are inherited directly from the many-electron quantum state, which in most practical cases has even a fixed particle number and magnetization. As it is proven in the Appendix \ref{sec:syminherit}, the third symmetry is in particular valid whenever the many-electron state is a singlet. Since the vast majority of molecular states in nature are singlets (otherwise the system would react with another one to form a closed-shell singlet structure), this is a reasonable assumption as well. Yet, for systems studied in solid state physics, this is not always the case and we therefore present in Appendix \ref{sec:general} a respective closed formula for the context of lattice models without assuming $[\rho, \hat{\vec{S}}^2]=0$.

In general, local symmetries of $\rho$ can be exploited to simplify the search space $\mathcal{D}_{sep}$ in \eqref{eqn:rel_ent} \cite{vollbrecht2001entanglement}.
To explain this, assume $[U(g),\rho]=0$ for all $g$ in a discrete group $G$, where $U(g)$ denotes its unitary representation on the Hilbert space. Any arbitrary state $\sigma$ can be turned into a $G$-symmetric one by applying the corresponding ``twirl'' $T_G(\cdot)$,
\begin{equation}\label{eqn:twirl}
    T_G(\sigma) =\frac{1}{|G|} \sum_{g \in G} U(g) \sigma U(g)^\dagger.
\end{equation}
Then one observes \cite{vollbrecht2001entanglement}:
\begin{equation}\label{eqn:entropy}
    \begin{split}
        S(\rho||\sigma) &= \frac{1}{|G|} \sum_{g \in G} S(U(g) \rho U(g)^\dagger || U(g) \sigma^\ast U(g)^\dagger)
        \\
        &= \frac{1}{|G|} \sum_{g \in G} S(\rho || U(g) \sigma^\ast U(g)^\dagger)
        \\
        &\geq S(\rho||T_G(\sigma)).
    \end{split}
\end{equation}
In the first line, we used the unitary invariance of the relative entropy $S$, in the second line the symmetry of $\rho$ and in the last one the convexity of $S$.
Since $U(g)$ was assumed to be local, $U(g)\equiv U_{\!A}(g)\otimes U_{\!B}(g)$, $T_G(\sigma)$ remains separable for all $\sigma \in \mathcal{D}_{sep}$. Hence, application of \eqref{eqn:entropy} to a minimizer state $\sigma^\ast \in \mathcal{D}_{sep}$ of \eqref{eqn:rel_ent} yields  the desired result: Whenever $\rho$ has a local symmetry $G$, we can restrict the search space $\mathcal{D}_{sep}$ to all $G$-symmetric states $\sigma \in T_G(\mathcal{D}_{sep})$.

In our case, one local symmetry is the one corresponding to $\hat{S}^z$.
Indeed, since $\hat{S}^z=\hat{S}^z_{\!A}\otimes \mathbb{1}_B+\mathbb{1}_A \otimes \hat{S}^z_{\!B}$ we have
\begin{equation}
U_{\hat{S}^z}(\alpha)\equiv \exp(i \alpha \hat{S}^z)= \exp(i \alpha \hat{S}_{A}^z) \otimes \exp(i \alpha \hat{S}_{B}^z),
\end{equation}
for all $\alpha$. Hence, by generalizing \eqref{eqn:twirl}, \eqref{eqn:entropy} to integrals \cite{vollbrecht2001entanglement}, we can restrict $\mathcal{D}_{sep}$ to states
\begin{eqnarray}
T_{\hat{S}^z}(\sigma)&=&\frac{1}{2\pi} \int_{0}^{2 \pi}\! \mathrm{d} \alpha \exp(i \alpha \hat{S}^z) \sigma \exp(-i \alpha \hat{S}^z) \nonumber \\
&=&\sum_{S^z} P_{S^z} \, \sigma \, P_{S^z},
\end{eqnarray}
which are block-diagonal with respect to the magnetization $S^z$, where $P_{S^z}$ denotes the projection onto the $S^z$-sector.
The same reasoning applies to the particle number operator $\hat{N}$. Also the particle number superselection rule \eqref{eqn:tilde} can be interpreted as a two-fold symmetry with corresponding $\hat{U}(\alpha)\equiv \exp{(i \alpha \hat{N}_{\!A})} \otimes \mathbb{1}_B$ and $\hat{U}(\beta)\equiv \mathbb{1}_A \otimes \exp{(i \beta \hat{N}_{\!B})}$, respectively.
\begin{table}[t!]
\begin{tabular}{|c|c|c|c|l|}
\hline
$\,\,N\,\,$                & $S^z$                   & \,\,\,$|\vec{S}| \,\,\,$            & $(N_A,N_B)$    & \qquad \quad State                                                                                                                           \rule{0pt}{2.6ex}\rule[-1.2ex]{0pt}{0pt} \\ \hline
0                  & 0                       & 0                      & $(0,0)$ & $\,\,|\Psi_1\rangle =|0\rangle \otimes |0\rangle$                                                                         \rule{0pt}{2.6ex}\rule[-1.2ex]{0pt}{0pt}   \\ \hline
\multirow{4}{*}{1} & \multirow{2}{*}{$1/2$}  & \multirow{2}{*}{$1/2$} & $(0,1)$ & $\,\,|\Psi_2\rangle =|0\rangle \otimes |\!\uparrow\rangle$                                                                     \rule{0pt}{2.6ex}\rule[-1.2ex]{0pt}{0pt}
\\ \cline{4-5}
                   &                         &                        & $(1,0)$ & $\,\,|\Psi_3\rangle =|\!\uparrow\rangle \otimes |0\rangle$                                                                     \rule{0pt}{2.6ex}\rule[-1.2ex]{0pt}{0pt}
                   \\ \cline{2-5}
                   & \multirow{2}{*}{$-1/2$} & \multirow{2}{*}{$1/2$} & $(0,1)$ & $\,\,|\Psi_4\rangle =|0\rangle \otimes |\!\downarrow\rangle$                                                                   \rule{0pt}{2.6ex}\rule[-1.2ex]{0pt}{0pt}
                   \\ \cline{4-5}
                   &                         &                        & $(1,0)$ & $\,\,|\Psi_5\rangle = |\!\downarrow\rangle \otimes |0\rangle$                                                                  \rule{0pt}{2.6ex}\rule[-1.2ex]{0pt}{0pt}
                   \\ \hline
\multirow{6}{*}{2} & \multirow{4}{*}{$0$}    & \multirow{3}{*}{0}     & $(2,0)$ & $\,\,|\Psi_6\rangle =|\!\uparrow\downarrow\rangle \otimes |0\rangle$                                                              \rule{0pt}{2.6ex}\rule[-1.2ex]{0pt}{0pt}
\\ \cline{4-5}
                   &                         &                        & $(0,2)$ & $\,\,|\Psi_7\rangle =|0\rangle \otimes |\!\uparrow\downarrow\rangle$                                                             \rule{0pt}{2.6ex}\rule[-1.2ex]{0pt}{0pt}
                   \\ \cline{4-5}
                   &                         &                        & $(1,1)$ & $\,\,|\Psi_8\rangle =\frac{|\uparrow\rangle|\otimes|\downarrow\rangle - |\downarrow\rangle \otimes |\uparrow\rangle}{\sqrt{2}}$ \rule{0pt}{2.6ex}\rule[-1.2ex]{0pt}{0pt}
                   \\ \cline{3-5}
                   &                         & 1                      & $(1,1)$ & $\,\,|\Psi_9\rangle =\frac{|\uparrow\rangle|\otimes|\downarrow\rangle + |\downarrow\rangle \otimes |\uparrow\rangle}{\sqrt{2}}$ \rule{0pt}{2.6ex}\rule[-1.2ex]{0pt}{0pt}
                   \\ \cline{2-5}
                   & $1$                     & 1                      & $(1,1)$ & $\,\,|\Psi_{10}\rangle =|\!\uparrow\rangle \otimes |\!\uparrow\rangle$                                                                \rule{0pt}{2.6ex}\rule[-1.2ex]{0pt}{0pt}
                   \\ \cline{2-5}
                   & $-1$                    & 1                      & $(1,1)$ & $\,\,|\Psi_{11}\rangle =|\!\downarrow\rangle \otimes |\!\downarrow\rangle$                                                           \rule{0pt}{2.6ex}\rule[-1.2ex]{0pt}{0pt}
                   \\ \hline
\multirow{4}{*}{3} & \multirow{2}{*}{$1/2$}  & \multirow{2}{*}{$1/2$} & $(2,1)$ & $\,\,|\Psi_{12}\rangle =|\!\uparrow\downarrow\rangle \otimes |\!\uparrow\rangle$                                                     \rule{0pt}{2.6ex}\rule[-1.2ex]{0pt}{0pt}
\\ \cline{4-5}
                   &                         &                        & $(1,2)$ & $\,\,|\Psi_{13}\rangle =|\!\uparrow\rangle \otimes |\!\uparrow\downarrow\rangle$                                                      \rule{0pt}{2.6ex}\rule[-1.2ex]{0pt}{0pt}
                   \\ \cline{2-5}
                   & \multirow{2}{*}{$-1/2$} & \multirow{2}{*}{$1/2$} & $(2,1)$ & $\,\,|\Psi_{14}\rangle =|\!\uparrow\downarrow\rangle \otimes |\!\downarrow\rangle$                                                 \rule{0pt}{2.6ex}\rule[-1.2ex]{0pt}{0pt}
                   \\ \cline{4-5}
                   &                         &                        & $(1,2)$ & $\,\,|\Psi_{15}\rangle =|\!\downarrow\rangle \otimes |\!\uparrow\downarrow\rangle$                                                   \rule{0pt}{2.6ex}\rule[-1.2ex]{0pt}{0pt}
                   \\ \hline
4                  & $0$                     & $0$                    & $(2,2)$ & $\,\, |\Psi_{16}\rangle =|\!\uparrow\downarrow\rangle \otimes |\!\uparrow\downarrow\rangle$                                          \rule{0pt}{2.6ex}\rule[-1.2ex]{0pt}{0pt}
\\
\hline
\end{tabular}
\caption{Decomposition of the Fock space into symmetry sectors labeled by the quantum numbers $(N, S^z,|\vec{S}|,{N}_{\!A}, {N}_{\!B})$.} \label{tab:sym}
\end{table}
It is astonishing, however, that even the twirl $T_{|\hat{\vec{S}}|}$ with respect to the total spin does not violate the separability criterion, despite its non-local character. This key result of our work is proven in the Appendix \ref{sec:ent_sym}.
In summary, we can restrict $\mathcal{D}_{sep}$ in \eqref{eqn:rel_ent} to the states sharing the same symmetries as $\tilde{\rho}$. Their fully-symmetric eigenstates $\ket{\Psi_i}$ are presented in  Table \ref{tab:sym} together with their quantum numbers $(N, S^z,|\vec{S}|,{N}_{\!A}, {N}_{\!B})$.

Since we can restrict $\mathcal{D}_{sep}$ to the density operators $\sigma = \sum_{i=1}^{16} q_i \ket{\Psi_i}\!\bra{\Psi_i}$ with the same eigenbasis as $\tilde{\rho}=\sum_{i=1}^{16} p_i \ket{\Psi_i}\!\bra{\Psi_i}$, the quantum relative entropy in \eqref{eqn:rel_ent} simplifies to the Kullback-Leibler divergence \cite{kullback1951information}
\begin{equation}
D_\text{KL}(\vec{p}\,||\vec{q}\,)=\sum_i p_i\log(p_i/q_i).
\end{equation}
The problem of finding the closest separable state is then transformed to that of finding the coefficients $q_i$ that minimizes $D_\text{KL}(\vec{p}\,||\vec{q}\,)$. Yet, it also remains to find a compact description of the convex set of $\vec{q}$ corresponding to fully symmetric separable states $\sigma$. One can exploit for this the fact that most of the states $\ket{\Psi_i}$ in Table \ref{tab:sym} are separable. To be more specific, one can even show (see Appendix \ref{sec:opt}) that all the entanglement in $\rho$ is confined to the sector $M = \mathrm{Span} \{ |\Psi_8\rangle, |\Psi_9\rangle, |\Psi_{10}\rangle, |\Psi_{11}\rangle \}$, for which eventually the Peres-Horodecki separability criterion \cite{peres1996separability,horodecki1997separability} becomes necessary and sufficient. Realizing all these technical ideas and solving the remaining minimization problem (see Appendix \ref{sec:analytic_form} for more details) leads finally to our closed formula for the relative entropy of entanglement $E(\rho)$. It depends on two crucial parameters, $t\equiv \max\{p_8,p_9\}$ and $r\equiv \min\{p_8,p_9\}+p_{10}+p_{11}$, where $p_i=\bra{\Psi_i}\tilde{\rho} \ket{\Psi_i}=\bra{\Psi_i}\rho \ket{\Psi_i}$. If
$r \geq t$, the state $\rho$ is unentangled, $E(\rho) =0$, and otherwise it is entangled with
\begin{equation}
        E(\rho) = r \log\left(\frac{2r}{r+t}\right) + t \log\left(\frac{2t}{r+t}\right). \label{eqn:rel_ent_formula}
\end{equation}
A generalization of this key result of our work is presented in Appendix \ref{sec:general} for the less relevant cases in which $\rho$ does not emerge from a \emph{singlet} many-electron state.

\section{Applications}\label{sec:applic}
The prospects of having a closed formula for the electron entanglement at hand can hardly be overestimated.
We briefly present in the following three applications which shall serve the scientific community as an inspiration. At the same time, this also provides first evidence for the broad applicability and potential relevance of our key result in physics, chemistry and materials science.

\subsection{Free Fermions and Kohn's Locality Principle}

\begin{figure}[t]
\centering
\includegraphics[scale=0.29]{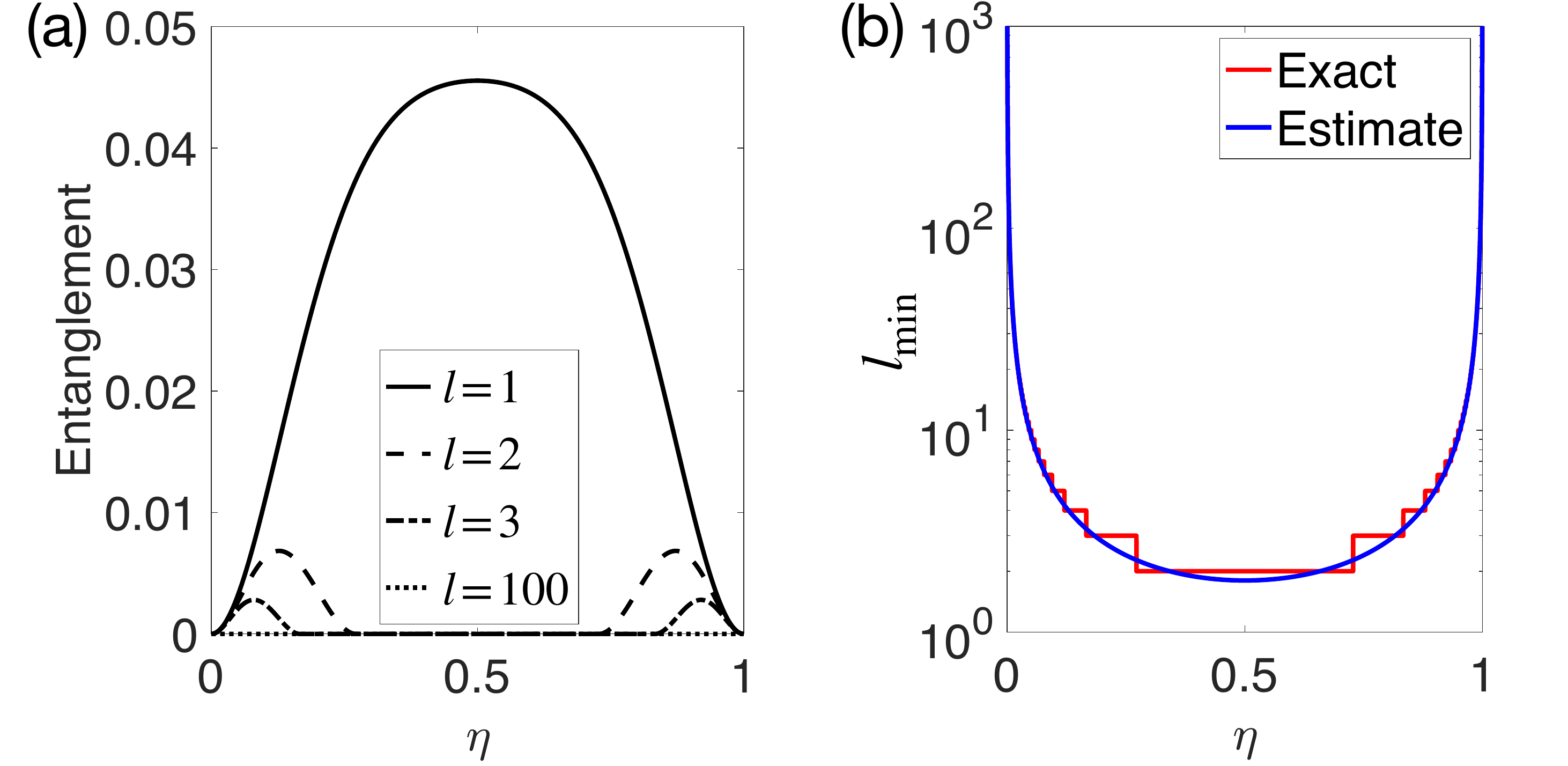}
\caption{Site-site entanglement in the ground state of free fermions on a 1D infinite lattice at various inter-site distances $l$ (a) and disentangling distance $l_\text{min}$ (b) as a function of the filling fraction $\eta$.}
\label{fig:FreeFermions}
\end{figure}
As a first example, we discuss free electrons on a 1D lattice with Hamiltonian
\begin{equation}\label{eqn:Hfree}
\hat{H}_\text{free} = -  \sum_{i,\sigma} (f^\dagger_{i\sigma} f^{\phantom{\dagger}}_{i+1,\sigma} + f^\dagger_{i+1,\sigma} f^{\phantom{\dagger}}_{i\sigma}),
\end{equation}
where $f_{i \sigma}^\dagger \,(f_{i \sigma})$ creates (annihilates) an electron with spin $\sigma$ at site $i$.
Since such quadratic models are analytically solvable, they have enjoyed a great popularity in the past decades in quantum many-body physics \cite{Peschel01,Peschel03,cheong2004free,peschel2009reduced,Calabrese_2016,vidmar2017free,eisler2020noncrit}. The $N$-electron ground state of \eqref{eqn:Hfree} follows as a Slater determinant with the $N$ energetically lowest spin-momentum states occupied and any physical quantity depends according to Wick's theorem on the corresponding one-particle reduced density operator only \cite{Gaudin60}.
By referring to this, we can provide (after a straightforward but lengthy calculation) a concise answer in \emph{analytic} terms to one central question: Does the ground state of \eqref{eqn:Hfree} exhibit long-distance entanglement?
For this, we consider the thermodynamic limit at filling factor $\eta$ and introduce the distance $l \equiv |i-j|$.
Result \eqref{eqn:rel_ent_formula} leads to a closed formula for the distance $l_{min}$ beyond which the entanglement between sites $i, j$ with $|i-j| \geq l_{min}$ vanishes,
\begin{equation}\label{eqn:dminleading}
l_\text{min} = \frac{\sqrt{2}}{\pi} \frac{1}{\eta(1-\eta)} + \mathcal{O}(1).
\end{equation}
Figure \ref{fig:FreeFermions} reveals that our leading order result \eqref{eqn:dminleading} approximates very well on the entire $\eta$-regime the numerically exact result for $l_{min}$. Most importantly, \eqref{eqn:dminleading} confirms in analytic terms that the site-site entanglement can become long-ranged, provided the electron ($\eta$) or hole filling factor ($1-\eta$) is sufficiently small.

At the same time, these results, particularly Eq.~\eqref{eqn:dminleading}, reveal univocally that the entanglement vanishes \emph{exactly} for most filling factors already when the sites are separated by just a few lattice constants. Accordingly, this resembles in the context of free electron chains Walter Kohn's seminal locality principle \cite{kohn1978locality,prodan2005nearsightedness}: ``\emph{[\ldots] local electronic properties, such as the density $n(r)$, depend significantly on the effective external potential only at nearby points.}'' \cite{prodan2005nearsightedness}. Indeed, this principle plays a pivotal role in modern chemistry since it ``\emph{[\ldots] can be viewed as underlying such important ideas as Pauling's ``chemical bond'', ``transferability'', and Yang's computational principle of ``divide and conquer''.}'' \cite{prodan2005nearsightedness}. It will therefore be one of the important future challenges to explore and quantify the long-distance entanglement in molecular systems based on our closed formula \eqref{eqn:rel_ent_formula}. If the entanglement turns out to vanish again exactly whenever two atomic orbitals are sufficiently far separated it will suggest a  refinement of the locality principle from a quantum information perspective. In that case, the quantum correlations are \emph{strictly} local in the sense that they do vanish \emph{exactly} beyond some critical separation distance. This then implies that the correlation function of any two local and spatially separated observables is merely of classical nature.

\subsection{Quantum Phase Transition}

\begin{figure}[t]
\centering
\includegraphics[scale=0.29]{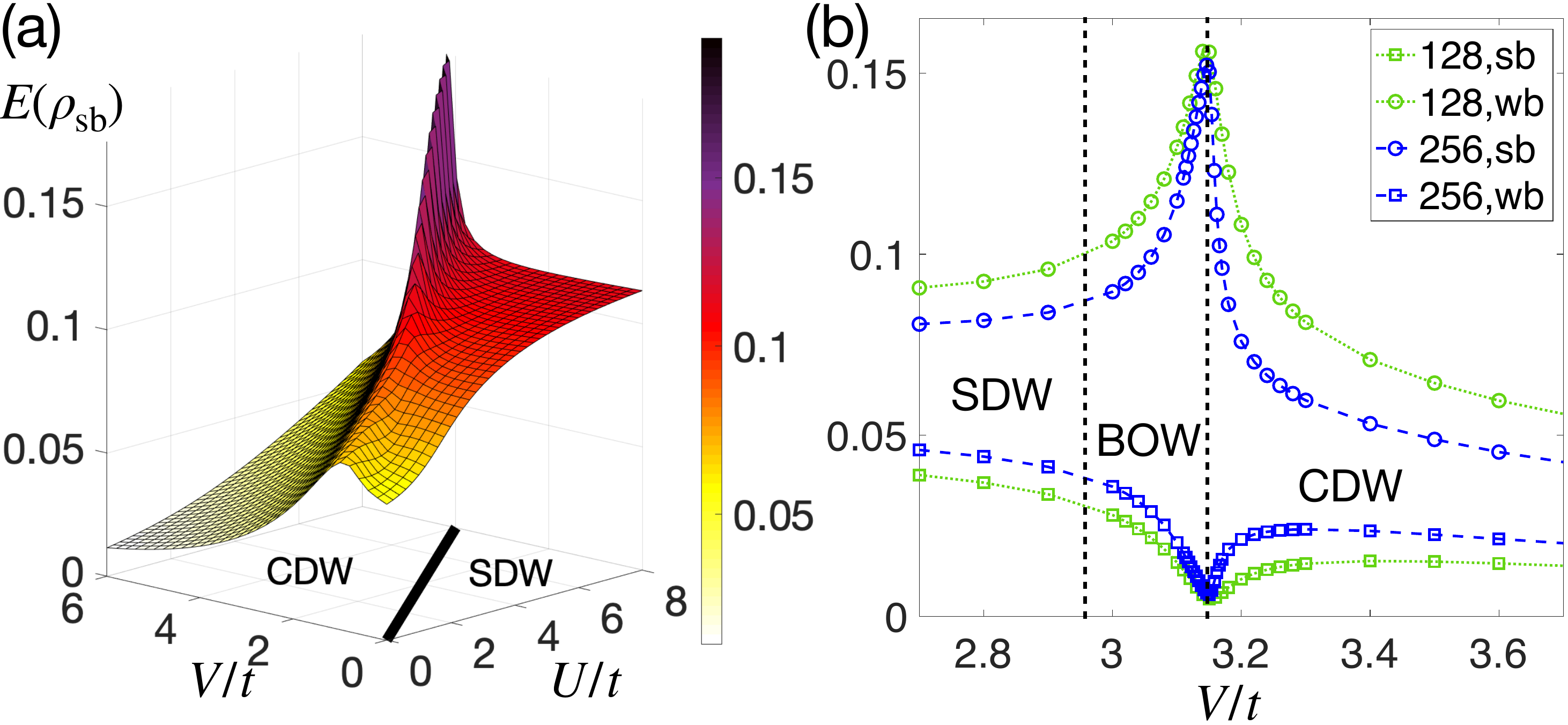}
\caption{(a) Nearest neighbor entanglement as a function of the on-site potential $U/t$ and nearest neighbor interaction strength $V/t$ with $50$ sites. (b) Entanglement in two adjacent pairs of nearest neighbors sharing a common site, $\rho_\text{sb}$ (strong bond) and $\rho_\text{wb}$ (weak bond) as function of $V/t$ at $U/t=6$, with $128$ and $256$ sites.}
\label{fig:EHM}
\end{figure}
The extended Hubbard model in 1D, described by the Hamiltonian
\begin{equation}\label{eqn:HamEHM}
\begin{split}
\hat{H} &= t \hat{H}_\text{free} + U \sum_i \hat{n}_{i\uparrow} \hat{n}_{i\downarrow} + V \sum_i \hat{n}_i \hat{n}_{i+1},
\end{split}
\end{equation}
where $\hat{n}_i\equiv \hat{n}_{i\uparrow}+\hat{n}_{i\downarrow}$, has a rich phase diagram at half filling, which has been extensively studied \cite{hirsch1984charge,cannon1991ground,voit1992phase,nakamura2000tricritical,tsuchiizu2002phase,sengupta2002bond,jeckelmann2002ground,gu2004entanglement,zhang2004dimerization,mund2009quantum}. Around $V \approx U/2$, between the well-understood charge density wave (CDW) and spin density wave (SDW) phase \cite{voit1992phase}, there exists an elusive, narrow bond order wave (BOW) phase \cite{nakamura2000tricritical}. Since the BOW phase is characterized  by nearest neighbor pairs that dimerize by forming alternatingly strong and weak bonds, this phase cannot be detected by the commonly used single-site entanglement entropy \cite{gu2004entanglement}. In striking contrast, our elaborated measure \eqref{eqn:rel_ent_formula} is ideally suited: By quantifying the entanglement between neighboring sites, we can not only detect the BOW phase and distinguish it form the CDW phase but also attribute its emergence to the three individual terms in the Hamiltonian \eqref{eqn:HamEHM}. The latter is due to the fact that each of them affects the two-orbital reduced density operator and \eqref{eqn:rel_ent_formula} in a different manner.
In Figure \ref{fig:EHM} (a) we present the nearest neighbor entanglement based on DMRG results for the case of $50$ sites. To also probe the bonding behaviour near $V = U/2$, we fix $U/t=6$ and present in the right panel the entanglement between the sites $i-1,i$ and between $i,i+1$, for $128$ and $256$ sites. The corresponding two-site reduced density operators  $\rho_\text{sb}$ and $\rho_\text{wb}$ describe the strong and weak bond, respectively. A peak (dip) is observed around $V/t \approx 3.17$ in the entanglement in $\rho_\text{sb}$ ($\rho_\text{wb}$), demonstrating the alternating bond strength of spontaneous dimerization. Therefore this critical value of $V/t$ is interpreted as a BOW-CDW transition point.
The transition SDW-BOW cannot be detected in the same manner since it is of infinite order \cite{nakamura2000tricritical}.

\subsection{Electronic Structure Analysis}

\begin{figure}[t!]
\centering
\includegraphics[scale=0.33]{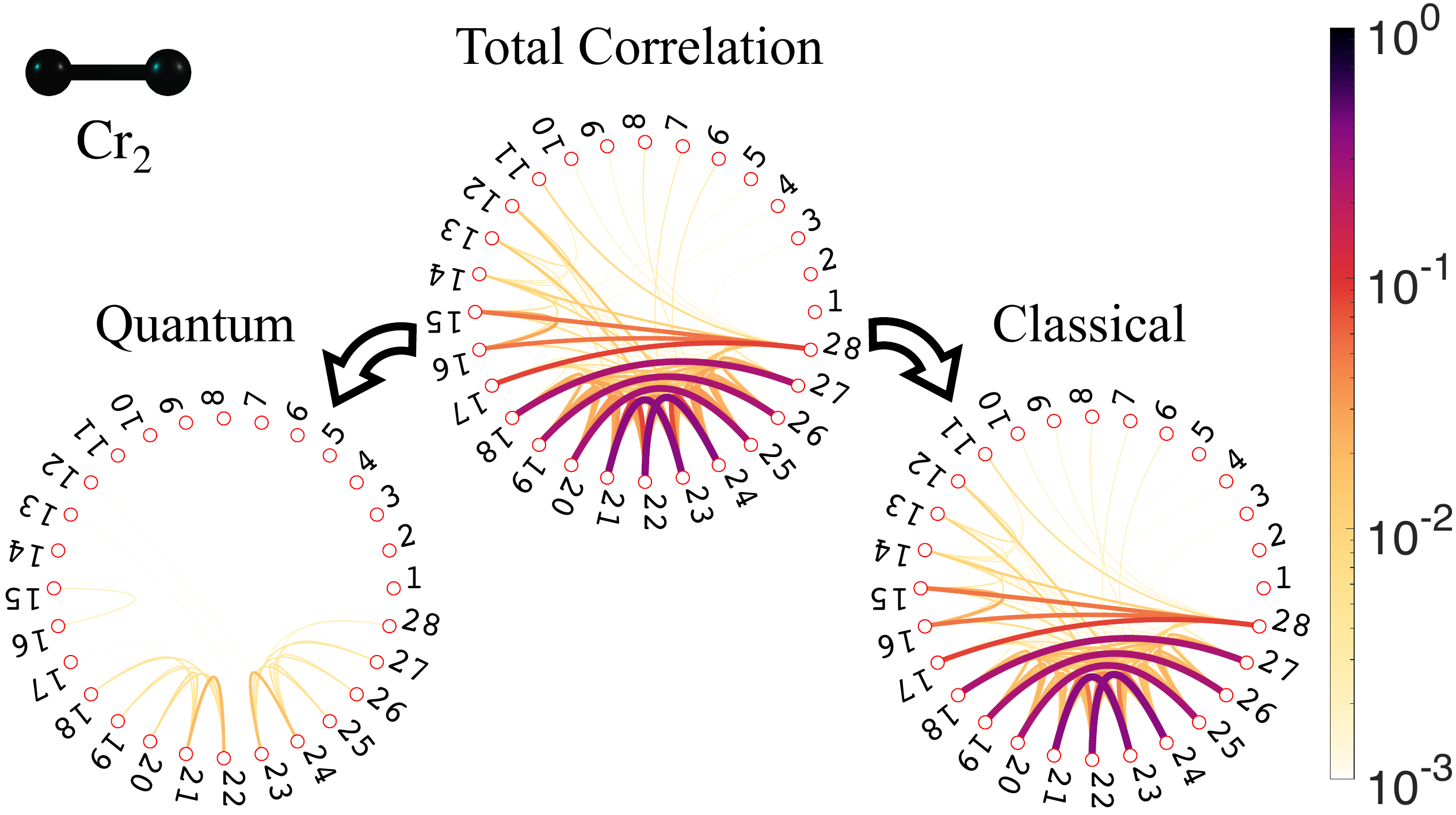}
\caption{Total correlation separated into entanglement (``Quantum'') and classical correlation (``Classical'') between 28 natural orbitals in the ground state of $\mathrm{Cr_2}$.}
\label{fig:qchem}
\end{figure}
As a third example, we discuss the electronic structure of a molecular system, with an emphasis on static correlation.
To calculate the entanglement and total correlation in the strongly correlated chromium dimer, we first determine its ground state $\ket{\Psi_{gs}}$ by a DMRG calculation involving 28 orbitals. For any two natural orbitals $\ket{\varphi_i},\ket{\varphi_j}$ (eigenstates of the orbital one-particle reduced density operator) we then determine the respective two-orbital reduced density operators
\begin{equation}
\rho_{i,j}= \mbox{Tr}_{\setminus \{i,j\}}[\ket{\Psi_{gs}}\!\bra{\Psi_{gs}}],
\end{equation}
as well as $\rho_{i/j}= \mbox{Tr}_{j/i}[\rho_{i,j}]$, by tracing out all the remaining orbitals.  The \emph{total correlation} between any two natural orbitals follows  as the quantum mutual information $S(\rho_{i,j}|| \rho_i \otimes \rho_j)$ \cite{vedral1997quantifying}. Remarkably, as it is revealed in Figure \ref{fig:qchem}, most of the correlation between the natural orbitals is classical, measured according to Ref.~\cite{Lindblad73,henderson2001classical} by $S(\sigma_{i,j}^\ast||\rho_i\otimes\rho_j)$, where $\sigma_{i,j}^\ast$ denotes the separable state closest to $\rho_{i,j}$. This surprising finding highlights the distinctive role of the natural orbitals as a reference basis for chemical computation. In this basis, the strong correlation contained in the molecule is rendered classical and the entangled is marginalized.

These findings also rationalize the ``zero seniority''-wavefunction ansatzes which play an important role since several decades in nuclear physics, with recent applications also in quantum chemistry for electronic structure calculations \cite{ring2004nuclear,stein2014seniority,boguslawski2017benchmark}. To explain this point, we recall that an $N$-electron wavefunction $|\Psi\rangle$ (with $N$ even) has zero seniority if its configuration interaction expansion with respect to the natural orbital basis simplifies
to
\begin{equation}\label{eqn:SZ}
|\Psi\rangle = \sum_{\vec{\nu}} c_{\vec{\nu}} |\vec{\nu}\,\rangle\,,
\end{equation}
where
\begin{equation}
|\vec{\nu}\,\rangle \equiv  |\nu_1,\nu_2,\ldots,\nu_d\rangle \equiv \prod_{i=1}^d(f_{i\uparrow}^\dagger f_{i\downarrow}^\dagger)^{\nu_i}\ket{0}\,,
\end{equation}
$\nu_i=0,1$ and $\sum_i \nu_i =N/2$.
This means that the expansion \eqref{eqn:SZ} contains only Slater determinants with \emph{paired} electrons. To understand what this means from a quantum information theoretical point of view, we determine for any two natural orbitals $i,j$ the respective two-orbital reduced density matrix. Its form follows directly as
\begin{equation}\label{eqn:SZrdm}
\rho_{i,j} = \sum_{\nu_i,\nu_j=0,1} \kappa_{i,j}^{(\nu_i,\nu_j)} \ket{\nu_i,\nu_j}\!\bra{\nu_i,\nu_j}\,
\end{equation}
with some non-negative coefficients $\kappa_{i,j}^{(\nu_i,\nu_j)}$ that depend on $\{c_{\vec{\nu}}\}$. The reduced states \eqref{eqn:SZrdm} are nothing else than classical mixtures of uncorrelated states $\ket{\nu_i,\nu_j}\!\bra{\nu_i,\nu_j} \equiv  \ket{\nu_i}\!\bra{\nu_i} \otimes \ket{\nu_j}\!\bra{\nu_j}$ and are thus unentangled. It is worth noticing that this would be not the case anymore if we expressed the state \eqref{eqn:SZ} with respect to a different orbital reference basis $\mathcal{B}$ (also since this would lead to unpaired electrons in the respective configuration interaction expansion).
Accordingly, the entanglement of a zero seniority wavefunction can be fully transformed into classical correlation through orbital rotations. This observation suggests to establish the sum
\begin{equation}\label{eqn:costf}
E^{(\mathcal{B})}(\rho) \equiv \sum_{1\leq i <j \leq d} E(\rho_{i,j}^{(\mathcal{B})})
\end{equation}
as an alternative measure of the effective seniority of an $N$-electron quantum state with respect to a chosen orbital reference basis $\mathcal{B}$. Minimizing then $E^{(\mathcal{B})}(\rho)$ with respect to all reference bases would yield the lowest possible effective seniority that could be reached through suitable orbital rotations. In that sense, the electron entanglement \eqref{eqn:rel_ent_formula} may serve through \eqref{eqn:costf} as a cost function for reducing the computational complexity of approaches to the ground state problem in quantum chemistry and nuclear physics.

\section{Summary and Conclusions}
Despite its comprehensive mathematical foundation, no closed formula has been known yet for measuring (faithfully) entanglement in generic many-electron quantum states. In the form of Eq.~\eqref{eqn:rel_ent_formula}, we succeeded in deriving such a formula for the entanglement between any two orbitals/sites in arbitrary many-electron quantum systems. For this, we exploited the common symmetries of realistic systems and incorporated the crucial particle number superselection rule. Because of the latter, our formula is operationally meaningful in contrast to most recent applications of quantum information theoretical tools in quantum many-body physics: It quantifies the true physical entanglement that could be extracted from a system and eventually be used as a resource for quantum information processing tasks.  For the sake of completeness, we also provide in Appendix \ref{sec:PSSR} the analog of \eqref{eqn:rel_ent_formula} for the weaker parity superselection rule.

The study of three physical examples has emphasized the potential significance of our key result \eqref{eqn:rel_ent_formula} and shall serve the scientific community as an inspiration for future applications. First, the presence of long-distant entanglement in free electron chains could be proven for low and high filling factors. For intermediate filling factors, however, the entanglement vanishes \emph{exactly} whenever the sites are separated by a few lattice constants. The latter refines from a quantum information theoretical perspective Kohn's locality principle which lies at the heart of modern quantum chemistry.  Second, the existence of the distinctive bond-order wave phase in the extended Hubbard model could be confirmed. Third, the total correlation in molecular systems was shown to be mainly classical and a distinctive pairing structure among natural orbitals could be revealed, even for the strongly correlated chromium dimer. In that sense, we could elucidate the success and the limitations of the zero-seniority wave function ansatz, as it is used in nuclear physics and quantum chemistry.  In the form of the averaged electron entanglement \eqref{eqn:costf} we put forth a corresponding measure of the effective seniority of a strongly correlated molecular ground state. Due to its concise quantum information theoretical character it may serve as a cost function, helping to improve the convergence rate of corresponding numerical implementations.

Last but not least, we recall that the two-orbital setting $\CC^4 \otimes \CC^4$ underlying our entanglement formula \eqref{eqn:rel_ent_formula} is information theoretical complete, in accordance with Coulson's important challenge \cite{Col00}: all relevant information about many-electron ground states is contained in the respective reduced density matrices. Indeed, in lattice models with hopping and interaction restricted to nearest neighbors, the calculation of the ground state energy involves effectively only the two-site reduced density matrices $\rho_{i,i+1}$. This and its generalization to continuous systems \cite{Col63,Mazz12,Mazz16} opens a novel avenue for advancing density functional theory since their functionals necessitate an ansatz of $\rho_{i,i+1}$ as function of the single-site density matrices. The latter means nothing else than quantifying various correlation types between sites (or localized orbitals), particularly the electron entanglement as described \emph{analytically} by Eq.~\eqref{eqn:rel_ent_formula}.

\begin{acknowledgments}
We thank S.\hspace{0.5mm}Mardazad for his support concerning the DMRG ground state calculations. We acknowledge financial support
from the Deutscher Akademischer Austauschdienst (DAAD Completion Scholarship 2020) (L.D.), the Deutsche Forschungsgemeinschaft (Grant SCHI 1476/1-1) (L.D., C.S.), the NKFIH through the Quantum Information National Laboratory of Hungary program, and Grants No. K124152, FK135220, K124351 (Z.Z.). The project/research is also part of the Munich Quantum Valley, which is supported by the Bavarian state government with funds from the Hightech Agenda Bayern Plus.
\end{acknowledgments}

\appendix

\section{Derivation of Entanglement Formulas} \label{sec:analytic_form}

In this section, we derive analytic formulas for the entanglement between two orbitals. In Section \ref{sec:ent_sym} we recap the role of local symmetry in simplifying the procedure for obtaining the closest separable state, and also provide a proof justifying the usage of the total spin symmetry despite it being a global symmetry. In Section \ref{sec:opt} we describe the procedure for calculating the closest separable state. In Section \ref{sec:NSSR} and Section \ref{sec:PSSR} we present the entanglement formulas in the presence of particle number and parity superselection rule, respectively.

We are interested in calculating the entanglement of a two-orbital (labelled $A$ and $B$) state $\rho$, with respect to the bipartition between the two orbitals $A$ and $B$, which refers to the splitting of the basis of the one-particle Hilbert space $\mathcal{H}^{(1)} = \mathcal{H}^{(A)} \oplus \mathcal{H}^{(B)}$ and the tensor product structure on the total Fock space $\mathcal{F}(\mathcal{H}^{(1)}) = \mathcal{F}(\mathcal{H}^{(A)}) \otimes \mathcal{F}(\mathcal{H}^{(B)})$. Separable states with respect to this bipartition are of the form
\begin{equation}
    \sigma = \sum_i p_i \sigma_A^{(i)} \otimes \sigma_B^{(i)}, \quad p_i \geq 0 \: \mathrm{and} \: \sum_i p_i = 1, \label{eqn:sep_states}
\end{equation}
where $\sigma^{(i)}_{A/B}$ are quantum states on orbital $A/B$. The relative entropy of entanglement\cite{vedral1997quantifying} of $\rho$ under the particle number superselection rule\cite{wick1970superselection} (N-SSR) is defined as the distance from the superselected state $\tilde{\rho}^\text{N}$
\begin{equation}
\rho \mapsto \tilde{\rho}^\text{N} = \sum_{N_{\!A}, N_{\!B} = 0}^2 P_{N_{\!A}}\! \otimes \! P_{N_{\!B}} \, \rho \, P_{N_{\!A}}\! \otimes\! P_{N_{\!B}}, \label{eqn:tilde}
\end{equation}
where $P_{N_{A/B}}$ is the projection onto the $N_{A/B}$-particle sector of the respective Fock space $\mathcal{F}_{A/B}$, to its closest separable state $\sigma^\ast$ measured by the relative entropy
\begin{equation}
    E(\rho) = \min_{\sigma \in \mathcal{D}_{sep}} S(\tilde{\rho}^\text{N}||\sigma) = S(\tilde{\rho}^\text{N}||\sigma^\ast).
\end{equation}
Here, $\mathcal{D}_{sep}$ is the set of separable states and the relative entropy $S(\rho||\sigma)$ is defined as
\begin{equation}
    S(\rho||\sigma) = \Tr[\rho(\log(\rho)-\log(\sigma))].
\end{equation}
We further assume the two-orbital state $\rho$ commutes with the total particle number $\hat{N}$, the total magnetization $\hat{S}^z$, and the total spin operator $\hat{\vec{S}}^2$. Notice that due to \eqref{eqn:tilde}, the superselected version of $\rho$ also commutes with the local particle number operators $\hat{N}_A \otimes \openone_B$ and $\openone_A \otimes \hat{N}_B$.

\subsection{Entanglement and Symmetry} \label{sec:ent_sym}

As explained in the main manuscript, an important insight from Ref.~\cite{vollbrecht2001entanglement} is that the closest separable state $\sigma^\ast$ also displays the same local symmetries as $\rho$, where the local symmetries are manifested as the invariance of $\rho$ and $\sigma$ under the twirl $T_G$ with respect to a group $G$ of local unitary operators defined as
\begin{equation}\label{eqn:twirl}
    T_G(\sigma) =\frac{1}{|G|} \sum_{g \in G} U(g) \sigma U(g)^\dagger.
\end{equation}
The effect of $T_G$ is demonstrated in Eq.~(5) in the main manuscript. To briefly summarize, if the group $G$ is generated by local observables, e.g. the magnetization $\hat{S}^z = \hat{S}^z_A \otimes \mathbbm{1}_B + \mathbbm{1}_A \otimes \hat{S}^z_B$, then $T_G(\sigma)$ is block diagonal with respect to the corresponding quantum numbers, in this case the magnetization $S^z$.

In Table I in the main manuscript we listed the simultaneous two-orbital eigenstates of $\hat{N}$, $ \hat{S}^z$, $ |\hat{\vec{S}}|$, $\hat{N}_A \otimes \mathbbm{1}_B$ and $\mathbbm{1}_A \otimes \hat{N}_B$, with eigenvalues $N$, $S^z$, $|\vec{S}|$, $N_A$ and $N_B$, respectively, which together generate the unitary group $G$. In this case all symmetry eigenspaces are one-dimensional, and a state that satisfies $\sigma = T_G(\sigma)$ can be written as
\begin{equation}
    \sigma = \sum_{i=1}^{16} q_i |\Psi_i\rangle\langle\Psi_i|, \quad q_i \geq 0, \: \sum_i^{16} q_i = 1, \label{eqn:coeff}
\end{equation}
where $q_i \equiv \langle \Psi_i| \sigma |\Psi_i\rangle$. The same applies to $\tilde{\rho}^\text{N}$ where the coefficients are denoted as $p_i \equiv \langle \Psi_i| \tilde{\rho}^\text{N} |\Psi_i\rangle = \langle \Psi_i| \rho |\Psi_i\rangle$ instead. Before continuing, we remark that the total spin $\hat{\vec{S}}^2$ symmetry also fits into this argument despite being a global symmetry.

\begin{lemm} \label{lemm:M}
If $\sigma = T_G(\sigma)$, then $\sigma$ is separable if and only if
\begin{equation}
    q_{10} q_{11} \geq \left( \frac{q_8-q_9}{2} \right)^2. \label{eqn:separability}
\end{equation}
\end{lemm}
\begin{proof}
Since $\sigma = T_G(\sigma)$, it can be expanded as \eqref{eqn:coeff}. We define the subspace $M = \mathrm{Span}\{|\psi_8\rangle, |\psi_9\rangle, |\psi_{10}\rangle, |\psi_{11}\rangle\}$. As $M$ is a two-qubit space, a state $\sigma$ with only support on $M$ is separable if and only if its partial transpose $\sigma^{T_B}$ defined via
\begin{equation}
    (\cdot)^{T_B} : |\psi_A\rangle\langle\phi_A| \otimes |\psi_B\rangle \langle \phi_B| \mapsto |\psi_A\rangle\langle\phi_A| \otimes |\phi_B\rangle \langle \psi_B| \label{eqn:PT}
\end{equation}
where $|\psi_{A/B}\rangle$ and $|\phi_{A/B}\rangle$ are the local basis vectors $\{|0\rangle, |\!\!\uparrow\rangle, |\!\!\downarrow\rangle, |\!\!\uparrow\!\downarrow\rangle\}$, only has non-negative eigenvalues, by the Peres-Horodecki criterion\cite{peres1996separability,horodecki1997separability}. In subspace $M$, this criterion translates to \eqref{eqn:separability}.

We now write $\sigma$ as $\sigma = c_1 \sigma|_M + c_2 \sigma|_{M^\perp}$, where $\sigma|_M$ ($\sigma|_{M^\perp}$) only has support on $M$ (the orthogonal subspace to $M$) and the positive numbers $c_1$ and $c_2$ sum to $1$. Notice that $\sigma|_{M^\perp}$ is always a mixture of product states, and hence separable. It then suffices to show that $\sigma$ is entangled if and only if $\sigma|_M$ is entangled. If $\sigma|_M$ is entangled, then its partial transpose necessarily contains at least one negative eigenvalue. As the map of partial transposition $(\cdot)^{T_B}$ is linear and $\sigma|_{M^\perp}$ is invariant under it, $\sigma^{T_B}$ also contains at least one negative eigenvalue. Then by the Peres-Horodecki criterion $\sigma$ is also entangled. On the other hand, if $\sigma|_M$ is separable, then $\sigma$ is also separable since it is a convex combination of two separable states.
\end{proof}

\begin{thrm}\label{prop:totalspin}
If $\sigma$ is separable and commutes with $\hat{N}_A\otimes \mathbbm{1}_B$, $\mathbbm{1}_A \otimes \hat{N}_B$ and $\hat{S}^z$, then $\tau_{|\hat{\vec{S}}|}(\sigma)$ is also separable where the twirl is with respect to the unitary group generated by $\vec{S}^2$.
\end{thrm}

\begin{proof}
We denote the group of unitaries generated by the operators $\hat{N}_A, \hat{N}_B$ and $\hat{S}^z$ as $G'$. According to Table I in the main manuscript, a state $\sigma$ that satisfies $\sigma = T_{G'}(\sigma)$ has the following form when restricted to the sector $M = \mathrm{Span}\{|\psi_8\rangle, |\psi_9\rangle, |\psi_{10}\rangle, |\psi_{11}\rangle\}$ (which is solely responsible for all the entanglement in $\sigma$, using the same argument in the proof of Lemma \ref{lemm:M})
\begin{equation}
    \sigma|_M = \begin{pmatrix}
        q_{10} & 0 & 0 & 0
        \\
        0 & q_8 & b & 0
        \\
        0 & \overline{b} & q_9 & 0
        \\
        0 & 0 & 0 & q_{11}
        \end{pmatrix},
\end{equation}
in the eigenbasis $|\Psi_i\rangle$. The quantities $q_i$ are defined as in \eqref{eqn:coeff} and $b = \langle \Psi_8 | \sigma |\Psi_9\rangle$. By the Peres-Horodecki criterion \cite{peres1996separability,horodecki1997separability}, the assumed separable state $\sigma$ satisfies
\begin{equation}
    \left(\frac{q_8-q_9}{2}\right)^2 + \mathrm{Im}(b)^2 \leq q_{10} q_{11}. \label{eqn:G}
\end{equation}
Now we apply $T_{|\hat{\vec{S}}|}$ to $\sigma$, i.e., twirling with respect to $|\hat{\vec{S}}|$, which is equivalent to the projection
\begin{equation}
    T_{|\hat{\vec{S}}|}(\cdot) = \sum_{|\vec{S}|} P_{|\vec{S}|} (\cdot) P_{|\vec{S}|}.
\end{equation}
Then it follows that the coherence term $b$ between the singlet ($|\Psi_8\rangle$) and triplet ($|\Psi_9\rangle$) state vanishes
\begin{equation}
    T_{|\hat{\vec{S}}|} \circ T_G'(\sigma)|_M = \begin{pmatrix}
        q_{10} & 0 & 0 & 0
        \\
        0 & q_8 & 0 & 0
        \\
        0 & 0 & q_9 & 0
        \\
        0 & 0 & 0 & q_{11}
        \end{pmatrix}. \label{eqn:sigma}
\end{equation}
Although the state \eqref{eqn:sigma} is diagonal, it may still contain entanglement, as the basis states $|\Psi_8\rangle$ and $|\Psi_9\rangle$ are entangled. Using again the Peres-Horodecki criterion, \eqref{eqn:sigma} is separable if and only if \eqref{eqn:separability} is met, which is satisfied specially by \eqref{eqn:G}. Therefore, if $\sigma= T_{G'}(\sigma)$ is separable, then $T_{|\hat{\vec{S}}|} \circ T_{G'}(\sigma)$ is also separable. Considering $G'$ is a \textit{local} unitary group, $T_G \equiv T_{|\hat{\vec{S}}|} \circ T_{G'}$ maps separable states to separable states.
\end{proof}

\subsection{Optimization} \label{sec:opt}

As $\Tilde{\rho}^\text{N}$ in \eqref{eqn:tilde} and any state $\sigma$ of the form \eqref{eqn:coeff} are diagonal in the same basis according to Table I in the main manuscript, the relative entropy $S(\tilde{\rho}^\text{N}||\sigma)$ can be written as
\begin{equation}
    S(\Tilde{\rho}^\text{N}||\sigma) = \sum_{i=1}^{16} p_i(\log(p_i)-\log(q_i)),
\end{equation}
where $p_i \equiv \langle \Psi_i | \tilde{\rho}^\text{N} |\Psi_i\rangle = \langle \Psi_i | \rho |\Psi_i\rangle$ and $q_i \equiv \langle \Psi_i | \sigma |\Psi_i\rangle$. Our goal is to minimize $S(\Tilde{\rho}^\text{N}||\sigma)$ by varying $\vec{q}=(q_1,q_2,\ldots,q_{16})$ under the constraint \eqref{eqn:separability}. The optimal set of coefficients $\{q^\ast_i\}_{i=1}^{16}$ then characterizes the closest separable state $\sigma^\ast$ to $\rho$.

Due to the inequality constraint \eqref{eqn:separability}, the optimization procedure is separated into two parts. First, we check if the global (without the constraint \eqref{eqn:separability}) solution satisfies already the separability condition \eqref{eqn:separability}. Because the closest state to $\tilde{\rho}^\text{N}$ measured by the relative entropy without any constraint is $\tilde{\rho}^\text{N}$ itself ($S(\rho||\rho)=0$), this step simply means that one first checks the separability of the state $\tilde{\rho}^\text{N}$. If $\tilde{\rho}^\text{N}$ is separable, then $E(\rho)=0$. Second, if $\tilde{\rho}^\text{N}$ is entangled, then one looks for the closest separable state that saturates the constraint \eqref{eqn:separability}, namely under the condition
\begin{equation}
    q_{10} q_{11} = \left( \frac{q_8-q_9}{2} \right)^2.
\end{equation}
In the Lagrange multiplier formalism, this is equivalent to minimizing the function
\begin{equation}
        \begin{split}
        F(\vec{q}\,) &= -\sum_{i}^{16} p_i \log(q_i) + \lambda \left[ \sum_{i=1}^{16} (q_i - p_i) \right]
        \\ &\quad + \mu \left[ q_{10}q_{11} - \left( \frac{q_8 - q_9}{2} \right)^2 \right] \label{eqn:minF}
    \end{split}
\end{equation}
with respect to $\vec{q}$, $\lambda$ and $\mu$, where the $\lambda$ ($\mu$) term encodes the normalization (separability) condition for $\sigma$. We focus on the most complicated case where the matrix $\tilde{\rho}^\text{N}$ is full rank, i.e., $p_i > 0$ for $i = 1,2,\ldots,16$. Before continuing with the derivation, we present a useful result.

\begin{thrm} \label{thrm:trace}
Let $\hat{Q}$ be an observable acting on $\mathcal{H}_A \otimes \mathcal{H}_B$ which can be written in the form $\hat{Q} = \hat{Q}_A \otimes \mathbbm{1}_B + \mathbbm{1}_A \otimes \hat{Q}_B$  and  has only two eigenvalues with corresponding eigenspaces $M_{1}$ and $M_{2}$.
Consider states of the form $\rho = a \rho_1 + (1-a) \rho_2$, where $\rho_{1/2}$ is a state with support only on $M_{1/2}$. Suppose any such state is separable if and only if both $\rho_1$ and $\rho_2$ are separable, then the closest separable state $\sigma^\ast$ to such a $\rho$ is of the form $\sigma^\ast = a \sigma^\ast_1 + (1-a) \sigma^\ast_2$, where $\sigma^\ast_{1/2}$ is a separable state with support only on $M_{1/2}$.

\end{thrm}

\begin{proof}
    From the assumption we know $\rho = T_{Q}(\rho)$ where $T_Q$ is the twirl with respect to the group of unitary operators generated by $\hat{Q}$. Then using the insight from Ref.\cite{vollbrecht2001entanglement}, the closest separable state $\sigma^\ast$ to $\rho$ also satisfies $\sigma^\ast = T_Q(\sigma^\ast)$, and
    \begin{equation}
        \sigma^\ast = b \sigma^\ast_1 + (1-b) \sigma^\ast_2, \quad b \in [0,1],
    \end{equation}
    and from the assumption, $\sigma^\ast_{1/2}$ are both separable and only has support on $M_{1/2}$. Then the relative entropy can be written as
    \begin{equation}
    \begin{split}
        S(\rho||\sigma^\ast) &= a S(\rho_1||\sigma^\ast_1) + (1-a) S(\rho_2 ||\sigma^\ast_2)
        \\
        &\quad + a(\log(a)-\log(b))
        \\
        & \quad + (1-a)(\log(1-a)-\log(1-b)).
    \end{split}
    \end{equation}
    For fixed $\rho_{1/2}$ and $\sigma_{1/2}^\ast$, the minimum is obtained when $b=a$.
\end{proof}

Theorem \ref{thrm:trace} can be generalized to the case with more than two eigenspaces. A direct consequence is, if we wish to find the closest separable state $\sigma^\ast$ to a state $\rho$ that is block diagonal, and these blocks are not coupled to each other by any separability criteria, then minimization can be carried out independently in each subspace. In particular, $\rho = \sigma^\ast$ when restricted to one-dimensional subspaces. We denote the set of indices of the eigenstates in Table I in the main manuscript spanning the subspace $M$ as $\mathcal{I} \equiv \{8,9,10,11\}$. Then we can immediately write down
\begin{equation}
    q_i^\ast = p_i, \quad i \notin \mathcal{I}, \label{eqn:Mperp}
\end{equation}
without performing any calculations. After fixing the coefficients \eqref{eqn:Mperp}, we are left with a new minimizing function which only concerns subspace $M$
\begin{equation}
        \begin{split}
        F(\vec{q}\,) &= \sum_{i \in \mathcal{I}} \left[ -p_i \log(q_i) + \lambda  (q_i - p_i) \right]
        \\
        & \quad +\mu \left[ q_{10}q_{11} - \left( \frac{q_8 - q_9}{2} \right)^2 \right]. \label{eqn:minF2}
    \end{split}
\end{equation}
Minimization of $F$ with respect to ${q_i}$ can be done by use of Mathematica. This leads to the results \eqref{eqns:soln} and \eqref{eqns:sol_gen} presented below.

\subsection{N-SSR Entanglement Formula} \label{sec:NSSR}

\subsubsection{The $p_{10} = p_{11}$ Case}\label{sec:p10p11}

We first look at a simpler situation when $p_{10} = p_{11}$. This additional condition is automatically satisfied if we assume an overall singlet on the total system that includes the two orbitals a subsystem, as we will explain in Section \ref{sec:syminherit}. In this case symmetry demands that $q_{10} = q_{11}$ and the quadratic constraint in \eqref{eqn:separability} reduces to linear ones. Assuming $p_8 > p_9$, the coefficients $\{q_i\}_{i \in \mathcal{I}}$ for the closest separable state in subspace $M$ are
\begin{equation}
    \begin{split}
        q_8^\ast &= \frac{r+t}{2}, \:\: q_9^\ast = \frac{r+t}{2r} p_9,\:\: q_{10}^\ast = q_{11}^\ast = \frac{r+t}{2r} p_{10},
        \label{eqns:soln}
    \end{split}
\end{equation}
where $t = \max\{ p_8, p_9 \}$ and $r = \min\{p_8,p_9\} + 2p_{10}$. For the case $p_9 > p_8$ one simply swaps $p_8 \leftrightarrow p_9$. This simple solution is due to the fact that the domain of search has linear boundaries and finitely many extreme points. The relative entropy of entanglement is
\begin{equation}
    \begin{split}
        E(\rho) = r \log\left(\frac{2r}{r+t}\right) + t\log\left(\frac{2t}{r+t}\right).\label{eqn:ent_form}
    \end{split}
\end{equation}

\subsubsection{The N-SSR General Case} \label{sec:general}

In the general case the solution is more involved. We assume $\rho$ displays a reflection symmetry between orbital $A$ and $B$, in replacement of the $\vec{S}^2$ symmetry. We again have the same expansion as in \eqref{eqn:coeff} but the condition $p_{10} = p_{11}$ is in general not met. Again, assuming $p_8 > p_9$,
\begin{equation}
    \begin{split}
        q_8^\ast &= \frac{A_1 + B_1 +\sqrt{C_1}}{4(s-p_9)},\quad q_9^\ast = \frac{A_1 - B_1 - \sqrt{C_1}}{4(s-p_8)},
        \\
        q_{10}^\ast &= p_{10} + \frac{p_8+p_9-q_8^\ast-q_9^\ast}{2},
        \\
        q_{11}^\ast &= p_{11} + \frac{p_8+p_9-q_8^\ast-q_9^\ast}{2}, \label{eqns:sol_gen}
    \end{split}
\end{equation}
where $A,B,C$ are polynomial functions of $p_8,p_9,p_{10},p_{11}$
\begin{equation}
\begin{split}
    A_1 &=  s^2 - (p_{10} - p_{11})^2, \quad
    B_1 = (p_8-p_9)s,
    \\
    C_1 &= (p_{10}+p_{11})^2(p_8 - p_9)^2 + 8 p_{10} p_{11} (2p_{10}p_{11}
    \\
    &\quad + (p_{10}+p_{11})(p_8 + p_9) + 2 p_8 p_9),
    \\
    s &= p_8 + p_9 + p_{10} + p_{11}.
\end{split}
\end{equation}
For the case $p_9 > p_8$ one again swaps $p_8 \leftrightarrow p_9$. One can check that when $p_{10} = p_{11}$, \eqref{eqns:sol_gen} reduces to \eqref{eqns:soln}. The N-SSR entanglement is calculated explicitly as
\begin{equation}
    E(\rho) = \sum_{i=1}^{16} p_i \log\left(\frac{p_i}{q_i^\ast}\right) = \sum_{i \in \mathcal{I}} p_i \log\left(\frac{p_i}{q_i^\ast}\right). \label{eqn:N_soln_gen}
\end{equation}

\subsection{P-SSR Entanglement Formula} \label{sec:PSSR}

If the N-SSR restriction is relaxed to the \textit{parity} superselection rule (P-SSR), while keeping the reflection symmetry between orbital $A$ and $B$ from Section \ref{sec:general}, an analytic formula for the site-site entanglement can still be obtained. In this case, the P-SSR restricted entanglement in $\rho$ is quantified as that in $\tilde{\rho}^\text{P}$\cite{bartlett2003entanglement}
\begin{equation}
    E(\rho) = \min_{\sigma\in \mathcal{D}_{sep}} S(\tilde{\rho}^\text{P}||\sigma).
\end{equation}
where
\begin{equation}
    \tilde{\rho}^\text{P} = \sum_{\tau, \tau' = \text{odd}, \text{even}} P_\tau \otimes P_{\tau'} \, \rho\, P_\tau \otimes P_{\tau'},
\end{equation}
and $P_{\text{even}}$ and $P_{\text{odd}}$ are the projections to the local even and odd parity subspaces, respectively.
Following a reasoning similar to the N-SSR case, a two-orbital state $\sigma$ that shares with $\tilde{\rho}^\text{P}$ the particle number, magnetization, reflection and local parity symmetry can be expanded as \eqref{eqn:coeff} with the additional changes to the eigenstates
\begin{equation}
    \Psi_{6/7} \longrightarrow \frac{|\Omega\rangle \otimes |\!\uparrow\downarrow\rangle \mp |\!\uparrow\downarrow\rangle \otimes |\Omega\rangle}{\sqrt{2}} \label{eqn:replace}
\end{equation}
in Table I in the main manuscript. Similar to the argument in the proof of Lemma \ref{lemm:M}, $\sigma$ is separable if and only if both $\sigma|_M$ and $\sigma|_{M'}$ are separable, where the subspace $M'$ is defined as $M'=\text{Span}\{|\Psi_1\rangle, |\Psi_6\rangle, |\Psi_7\rangle, |\Psi_{16}\rangle\}$. Then by Theorem \ref{thrm:trace}, the optimization for obtaining the closest separable state $\sigma^ast$ can be carried out independently in the subspaces $M$ and $M'$. The optimization in the sector $M$ is already covered in Section \ref{sec:NSSR}. As for the sector $M'$, using again the Peres-Horodecki criterion\cite{peres1996separability,horodecki1997separability}, $\sigma|_{M'}$ is entangled if and only if
\begin{equation}
    q_1 q_{16} < \left( \frac{q_6 - q_7}{2} \right)^2.
\end{equation}

The minimization scheme in sector $M'$ likewise splits into two parts. First we check the separability of $\tilde{\rho}^\text{P}|_{M'}$. If $\tilde{\rho}^\text{P}|_{M'}$ is separable, then $E(\rho)$ is the same as the N-SSR entanglement, which can be calculated according to Section \ref{sec:NSSR}. If $\tilde{\rho}^\text{P}|_{M'}$ is entangled, then (when $p_{1} = p_{16}$ in analogy to Section \ref{sec:p10p11}, which is met, for example, by a particle-hole symmetrized $\rho$)
\begin{equation}
\begin{split}
E(\rho) &= r \log\left(\frac{2r}{r+t}\right) + t\log\left(\frac{2t}{r+t}\right)
\\
& +  r' \log\left(\frac{2r'}{r'+t'}\right) + t'\log\left(\frac{2t'}{r'+t'}\right), \label{eqn:rel_ent_formula2}
\end{split}
\end{equation}
where  $t' = \max\{p_6,p_7\}$ and $r' = \min\{p_6,p_7\} + p_1 + p_{16}$. The quantities $r$ and $t$ are the same as in \eqref{eqn:ent_form}. In case the particle-hole symmetry is lifted ($p_1 \neq p_{16}$), the expansion coefficients for the closest separable state in the relevant $M'$ sector $\sigma|_{M'}$ are (for $p_6 \geq p_7$)
\begin{equation}
    \begin{split}
        q_6^\ast &= \frac{A_2 + B_2 +\sqrt{C_2}}{4(s'-p_7)},\quad q_7^\ast = \frac{A_2 - B_2 - \sqrt{C_2}}{4(s'-p_6)},
        \\
        q_{1}^\ast &= p_{1} + \frac{p_6+p_7-q_6^\ast-q_7^\ast}{2},
        \\
        q_{16}^\ast &= p_{16} + \frac{p_6+p_7-q_6^\ast-q_7^\ast}{2},
    \end{split}
\end{equation}
where $A_2,B_2,C_2$ are polynomial functions of $p_i$ for $i\in \mathcal{I}'\equiv \{1,6,7,16\}$ (recalling that $p_6$ and $p_7$ are modified according to \eqref{eqn:replace}),
\begin{equation}
\begin{split}
    A_2 &=  s'^2 - (p_{1} - p_{16})^2, \quad
    B_2 = (p_6-p_7)s,
    \\
    C_2 &= (p_{1}+p_{16})^2(p_6 - p_7)^2 + 8 p_{1} p_{16} (2p_{1}p_{16}
    \\
    &\quad + (p_{1}+p_{16})(p_6 + p_7) + 2 p_6 p_7),
    \\
    s' &= p_1 + p_6 + p_{7} + p_{16}.
\end{split}
\end{equation}
The P-SSR entanglement is calculated explicitly as
\begin{equation}
    E(\rho) = \sum_{i=1}^{16} p_i \log\left(\frac{p_i}{q_i^\ast}\right) = \sum_{i\in \mathcal{I},\mathcal{I}'} p_i \log\left(\frac{p_i}{q_i^\ast}\right). \label{eqn:P_soln_gen}
\end{equation}

To summarize the applicable scenarios for the formulas \eqref{eqn:ent_form}, \eqref{eqn:N_soln_gen}, \eqref{eqn:rel_ent_formula2} and \eqref{eqn:P_soln_gen}, we tabulated the necessary symmetries in Table \ref{tab:sym_required}.

\begin{table}[t!]
\begin{tabular}{|r|c|c|}
\hline
\multicolumn{1}{|r|}{\multirow{3}{*}{N-SSR}} & \eqref{eqn:ent_form}                                                                    & \eqref{eqn:N_soln_gen} \rule{0pt}{2.6ex}\rule[-1.2ex]{0pt}{0pt}
\\
\cline{2-3}
\multicolumn{1}{|r|}{}                       & \multirow{2}{*}{\begin{tabular}[c]{@{}c@{}}$S^z$, global singlet\end{tabular}}             & \multirow{2}{*}{$S^z$, $\vec{S}^2$/reflection}
\rule{0pt}{2.6ex}\rule[-1.2ex]{0pt}{0pt}
\\
\multicolumn{1}{|r|}{}                       &                                                                                                           &                                           \rule{0pt}{2.6ex}\rule[-1.2ex]{0pt}{0pt}
\\ \hline
\multirow{3}{*}{P-SSR}                       & \eqref{eqn:rel_ent_formula2}                                                           & \eqref{eqn:P_soln_gen}
\rule{0pt}{2.6ex}\rule[-1.2ex]{0pt}{0pt}
\\ \cline{2-3}
                                             & \multirow{2}{*}{\begin{tabular}[c]{@{}c@{}}$N$, $S^z$,
                                             \\ particle-hole\end{tabular}} & \multirow{2}{*}{$N$, $S^z$, reflection}
                                             \rule{0pt}{2.6ex}\rule[-1.2ex]{0pt}{0pt}
                                             \\
                                             &                                                                                                           &
                                             \\ \hline
\end{tabular}
\caption{Required symmetries on the two-orbital state $\rho$ for using formulas \eqref{eqn:ent_form}, \eqref{eqn:N_soln_gen}, \eqref{eqn:rel_ent_formula2} and \eqref{eqn:P_soln_gen}. $N$, $S^z$ and $\vec{S}^2$ refer to the total particle number, magnetization and total spin symmetry of the two-orbital state, respectively. The reflection symmetry is with respect to the reflection between orbital $A$ and $B$. The global singlet refers to the situation where state $|\Psi\rangle$ on the total system, including orbitals $A$ and $B$ as a subsystems, is a singlet. The particle-hole symmetry condition on $\rho$ can also be replaced by a symmetry that realize the equality $p_1 = p_{16}$.}
\label{tab:sym_required}
\end{table}

\section{Symmetry Inheriting} \label{sec:syminherit}

In this section, we treat the two orbitals as a subsystem of a larger set of orbitals, and discuss how are local and global symmetries in the total system inherited by the two-orbital subsystem.

With respect to a bipartition $I:J$ of a set of orbitals, there are two types of symmetry, local and global. Local symmetries are associated with conserved observables that take the form
\begin{equation}
 \hat{Q} =  \hat{Q}_I \otimes \openone_J + \openone_I \otimes  \hat{Q}_J. \label{eqn:local_symm}
\end{equation}
The group of unitary operators generated by $ \hat{Q}$ is therefore also local in the sense that its elements are factorized, i.e. $U \equiv \exp(i\alpha \hat{Q}) = U_I \otimes U_J$ where $U_{I/J} = \exp(i\alpha  \hat{Q}_{I/J})$. If the quantity $Q$ is conserved on the total system, namely the total state $\rho_\text{tot}$ satisfies $U \rho_\text{tot} U^\dagger = \rho_\text{tot}$, then the quantity $Q_I$ is also conserved in subsystem $I$ manifested as $U_I \rho_I U_I^\dagger= \rho_I$ which follows directly from the local unitary invariance of the partial trace. In short, local symmetries of the total state are naturally inherited by the reduced states, as expected.

The other type of symmetries is associated with observables that cannot be cast into the form \eqref{eqn:local_symm}. These symmetries are global, and in general not inherited by the reduced states on a subsystem. However, we argue in Theorem \ref{thrm:symminherit} that if we further assume that the total state $|\Psi\rangle$ is a pure singlet, then the reduced state on any subsystem also commutes with the total spin operator.

\begin{thrm}\label{thrm:symminherit}
Let $I$ and $J$ be two subsystems corresponding to a bipartition of orbitals. If the total state $|\Psi\rangle$  of the system  is a singlet state, namely $\hat{\vec{S}}^2|\Psi\rangle = \hat{S}^z |\Psi\rangle = 0$, then the reduced state $\rho_I = \Tr_J(|\Psi\rangle \langle\Psi|)$ satisfies $\left[\rho_I,  \hat{\vec{S}}_I^2\right]  =\left[\rho_{I},  \hat{S}^z_I\right] = 0$.
\end{thrm}
\begin{proof}
We know that $\rho_\text{tot} = |\Psi\rangle \langle \Psi|$ commutes with both $\hat{\vec{S}}^2$ and $ {S}^z$. For the commutator with the magnetization of subsystem $I$, we take the partial trace of the following commutator,
\begin{equation}
\begin{split}
        0 &= \Tr_J\left( \left[ \rho_\text{tot},  \hat{S}^z\right] \right) = \Tr_J \left( \left[ \rho_\text{tot},  \hat{S}^z_I +  \hat{S}^z_J \right] \right)
        \\
        &= \Tr_J \left( \left[ \rho_\text{tot},  \hat{S}^z_I \right] \right) = \left[ \rho_I,  \hat{S}^z_I \right]. \label{eq:Sz}
\end{split}
\end{equation}
$\Tr_J \left( \left[ \rho_\text{tot},  \hat{S}^z_J \right] \right) = 0$ due to cyclicity of the partial trace. For the commutator with the total spin of subsystem $I$, we look at the following quantity
\begin{equation}
\begin{split}
\Tr_J \left( \left[\rho_\text{tot}, \hat{\vec{S}}^2\right] \right) &= \Tr_J \left( \left[ \rho_\text{tot}, \hat{\vec{S}}_I^2 \right] \right) + \Tr_J \left( \left[ \rho_\text{tot}, \hat{\vec{S}}_J^2 \right] \right)
\\
& \quad + 2\Tr_J \left( \left[ \rho_\text{tot}, \hat{\vec{S}}_I \cdot \hat{\vec{S}}_J \right] \right).
\end{split}
\end{equation}
The first term is the sought after $\left[\rho_I, \vec{S}_I^2\right]$. The second term vanishes again due to the cyclicity of the partial trace. Then we are left with the last term. We rewrite it as
\begin{equation}
    \begin{split}
        &\quad \: \, \Tr_J \left( \left[ \rho_\text{tot}, \hat{\vec{S}}_I \cdot \hat{\vec{S}}_J \right] \right)
        \\
        &= \Tr_J \left( \left[ \rho_\text{tot}, \hat{S}^x_I \hat{S}^x_J + \hat{S}^y_I \hat{S}^y_J + \hat{S}^z_I \hat{S}^z_J \right] \right). \label{eqn:crossterms}
    \end{split}
\end{equation}
We us the fact that $|\Psi\rangle$ is an eigenstate of $\hat{S}^z$ with eigenvalue $0$ and write down its Schmidt decomposition
\begin{equation}
    |\Psi\rangle = \sum_{i} \lambda_i  |s_i;a_i \rangle_I \otimes |\!-\!s_i;b_i \rangle_J. \label{eqn:psi}
\end{equation}
$a_i$ and $b_i$ denotes degrees of freedom within the degeneracy classes (e.g., arrangements of spin up and spin down electrons). Then
\begin{equation}
    \begin{split}
        \Tr_J \left( \rho_\text{tot} \hat{S}^z_J\right) = |\lambda_i|^2 ( - s_i) |s_i;a_i\rangle\langle s_i;a_i|,
    \end{split}
\end{equation}
which commutes with $ \hat{S}^z_I$. Notice for $\rho_\text{tot} = |\Psi\rangle \langle \Psi|$ of the form \eqref{eqn:psi} we have
\begin{equation}
    \Tr_J \left(\left[ \rho_\text{tot}, \hat{S}^z_I \hat{S}^z_J\right] \right) = \left[ \hat{S}^z_I, \Tr_J \left( \rho_\text{tot} \hat{S}^z_J\right) \right],
\end{equation}
from which deduce that
\begin{equation}
    \Tr_J \left( \left[ \rho_\text{tot}, \hat{S}^z_I \hat{S}^z_J\right] \right) = 0.
\end{equation}
The $x$- and $y$-component spin operator terms in \eqref{eqn:crossterms} vanishes by the same argument, since singlets states are rotationally invariant.
\end{proof}
We remark that the proof for Theorem \ref{thrm:symminherit} can also be extended to mixed singlets, by the linearity of the partial trace and commutation. With this additional assumption that $\rho_\text{tot}$ is a singlet (pure or mixed), the two-orbital reduced state $\rho$ on orbitals $A$ and $B$ also enjoys the spin-flip symmetry, manifested as (assuming the total system has $K$ orbitals)
\begin{eqnarray}
\lefteqn{\langle \uparrow, \uparrow\!| \Tr_{\setminus\{A,B\}}(\rho_\text{tot}) |\!\uparrow,\uparrow\rangle} \qquad \nonumber  \\
&= &\langle \uparrow, \uparrow| \Tr_{\setminus \{A,B\}}((U^\dagger)^{\otimes K}\rho_\text{tot} \, U^{\otimes K}) |\uparrow,\uparrow\rangle
\nonumber \\
&=& \langle \uparrow, \uparrow\!|(U^\dagger)^{\otimes 2} \rho \, U^{\otimes 2} |\!\uparrow,\uparrow\rangle
\nonumber \\
&= &\langle \downarrow, \downarrow\!|\rho |\!\downarrow,\downarrow\rangle,
\end{eqnarray}
where $U$ is a local basis transformation acting that maps $|\!\uparrow\rangle$ to $|\!\downarrow\rangle$ and vice versa. Referring to Table I in the main manuscript, this translates to the condition $p_{10} = p_{11}$ in Section \ref{sec:p10p11}, which allows us to use the simple formula \eqref{eqn:ent_form} to calculate the site-site entanglement under N-SSR.

\bibliography{EntOrb-v8.bib}

\end{document}